\documentclass[11pt]{article}

\usepackage[preprint]{acl}
\usepackage{booktabs}
\usepackage{pifont}
\usepackage{cleveref}
\Crefname{figure}{Fig.}{Figs.}
\Crefname{table}{Tab.}{Tabs.} 
\crefformat{section}{\S#2#1#3}
\usepackage{multirow}

\newcommand{\data}{{\sc LegalPincite}}
\newcommand{\cmark}{\ding{51}}
\newcommand{\xmark}{\ding{55}}

\usepackage{times}
\usepackage{latexsym}

\usepackage[T1]{fontenc}

\usepackage[utf8]{inputenc}

\usepackage{microtype}

\usepackage{inconsolata}

\usepackage{graphicx}

\title{LegalPincite: Multi-level Legal Information Retrieval Dataset}

\author{
 \textbf{Theresia Veronika Rampisela\textsuperscript{1,2}},
 \textbf{Henrik Palmer Olsen\textsuperscript{3}},
 \textbf{Giovanni Colavizza\textsuperscript{1,2}}
 \\
 \textsuperscript{1}Department of Communication, University of Copenhagen, Denmark \\
 \textsuperscript{2}Centre for Digital and Computational Humanities, University of Copenhagen, Denmark \\
 \textsuperscript{3}Faculty of Law, University of Copenhagen, Denmark
\\
 \small{
   \texttt{\{thra, colavizza\}@hum.ku.dk} \quad \texttt{henrik@jur.ku.dk}
 }
}

\begin{document}
\maketitle
\begin{abstract}
    A common task in legal Information Retrieval (IR) is to find relevant legal sources from case-law collections. While legal practice often requires pinpoint citations (pincites) to specific case paragraphs, most existing public legal IR datasets lack paragraph-level citation annotations. Yet, publicly available datasets with such information contain data leakage in the query text and exclude paragraphs that are neither citing nor cited from the corpora, creating an unrealistic and oversimplified retrieval setting, potentially leading to inflated performance. 
    To address these limitations, we contribute a large-scale legal IR dataset constructed from Court of Justice of the European Union (CJEU) judgments. The dataset  contains: 
    (i) masked case/paragraph queries, with removed citation information;
    (ii) a corpus that includes all paragraphs; and
    (iii) case- and paragraph-level ground truth citations, with partial human expert validation. 
    Our dataset supports both the  development and rigorous evaluation of legal IR methods, at multiple query-document levels (case-to-case, paragraph-to-case, and paragraph-to-paragraph retrieval). Link to dataset and code: \href{https://huggingface.co/datasets/theresiavr/legalpincite}{huggingface.co/datasets/theresiavr/legalpincite}.
\end{abstract}

\section{Introduction}

\begin{figure}
    \centering
    \includegraphics[width=\linewidth, trim=0.75cm 5.3cm 0.75cm 0cm, clip=True]{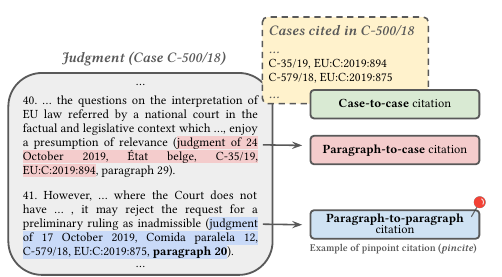}
    \caption{Citation types in our {\sc LegalPincite} dataset.}
    \label{fig:teaser}
\end{figure}

Legal Information Retrieval (IR) aims to identify the most suitable legal information with respect to an input query \cite{SANSONE2022101967}. An example of information that is usually searched for is \textit{relevant citations} to legal sources, which can be useful to support legal analysis \cite{hou-etal-2025-clerc}, substantiate judgments \cite{Shulayeva2017}, and communicate existing precedents, giving weight and authority to a decision for the considered case \cite{auslawcitebenchmark}. These citations can be made on a case level to another case (case-to-case), on a paragraph level to another case (paragraph-to-case), or on a paragraph-to-paragraph basis \cite{Panagis2017JURIX}. The latter is known as \textit{pincite} or pinpoint citations (see example in \Cref{fig:teaser}), and is a common legal practice especially in case law \cite{Olsen2026ProvidingPincite}.

\begin{table*}[t]
\centering
\resizebox{\textwidth}{!}{
\begin{tabular}{p{5.25cm} l c c c c c c}
\toprule
\multirow{2}{*}{\textbf{Dataset}} &
\multirow{2}{*}{\textbf{Source}} &
\multicolumn{3}{c}{\textbf{Query-Document Level}} &
\multirow{2}{*}{\textbf{Leakage Mitigation}} &
\multirow{2}{*}{\textbf{All Paragraphs}} &
\multirow{2}{*}{\textbf{Human Validation}} \\
\cmidrule(lr){3-5}
& & case-case & par-case & par-par \\
\midrule
COLIEE \cite{coliee} &  Federal Court of Canada & \cmark & \xmark & \xmark & \cmark & \xmark & \xmark  \\
AusLaw \cite{auslawcitebenchmark} & NSW Caselaw (Australia) & \cmark & \xmark  & \xmark  & \cmark & \cmark & \xmark  \\
IRLeD \cite{IRLeD2017} & Supreme Court of India & \cmark & \xmark & \xmark & \cmark & \cmark & \xmark \\
ECtHR-PCR \cite{t-y-s-s-etal-2024-ecthr} & ECtHR (Europe) & \cmark & \xmark & \xmark & \cmark & \cmark & \xmark \\
CLERC \cite{hou-etal-2025-clerc} & US federal case law & \xmark & \cmark & \xmark & \cmark & \cmark & \xmark \\
Caselaw \cite{USSC_CaseLaw,Locke2018CaseLaw} & US Supreme Court & \xmark & \cmark & \xmark & \cmark & \cmark$^\dagger$ & \cmark \\
\citet{Mori2026ICAIL} &  Court of Justice (EU) & \xmark & \xmark & \cmark & \xmark & \xmark & \xmark \\
LePaRD \cite{mahari-etal-2024-lepard} &  US federal case law & \xmark & \xmark & \cmark & \cmark$^\dagger$ & \xmark & \xmark \\
\citet{Olsen2023ReFraming} & Court of Justice (EU) & \xmark & \xmark & \cmark & \xmark & \xmark & \cmark \\
\citet{Olsen2026ProvidingPincite} & Court of Justice (EU) & \xmark & \xmark & \cmark & \xmark & \cmark$^\dagger$ & \cmark \\
\midrule
\data{} (ours) & Court of Justice (EU) & \cmark & \cmark & \cmark & \cmark & \cmark & \cmark \\
\bottomrule
\end{tabular}
}
\caption{Datasets for legal case retrieval or paragraph citation retrieval in English, based on query-document level, citation leakage mitigation on queries, all paragraphs (not only citing/cited ones), and (partial) human validation on ground-truth relevance. 
$\dagger$ means the component is not publicly available.
}
\label{tab:dataset_comparison}
\end{table*}

Case law pincites are important for practitioners as legal documents tend to be very long and may relate to several different legal issues \cite{Zhang2007SemanticNetwork,Geist2007LOAIT}. 
For instance, in the famous Bosman ruling (C-415/93), the judgment in the case addresses at least four legal issues: 
(i) the Court’s jurisdiction to hear a preliminary ruling (Bosman established a balanced approach between deference
to the national court and ensuring that the question asked bears a real relation to the actual facts of the case); (ii) the application of fundamental freedoms to private entities (expanding the horizontal effect of treaty law); (iii) defining
what constitutes an obstacle to a fundamental freedom and justification of such obstacles (expanding on the right to
free movement of labour); and (iv) the temporal effects of Court of Justice of the European Union (CJEU) decisions (narrowing the retroactive application of the
findings to prior transfer agreements). Each of these issues is relevant to its own legal issue in the overall EU law system. Hence, there is a need, when citing Bosman as a source of law (precedent) to specify which part of the Bosman ruling is relevant. 

The court itself, as well as other lawyers who engage with EU law, routinely indicates specific paragraphs when citing case law as legal authority. This illustrates how precedent is better understood as a sub case level phenomena. Thus, when seeking to understand what CJEU case law is important, being able to retrieve cited paragraphs is an important task. While non-cited case paragraphs could also be relevant, cited paragraphs provide evidence for what the court itself considers relevant to cite as authority for the decision in the citing case.

To our knowledge, only four datasets are available for paragraph-to-paragraph legal citation retrieval \cite{Olsen2023ReFraming,Olsen2026ProvidingPincite,Mori2026ICAIL,mahari-etal-2024-lepard} (see \Cref{tab:dataset_comparison}). These datasets suffer from two serious limitations. Firstly, none of them mask citation-related information in the queries, e.g., case title, cited paragraph number, or case-related entities. As such, the query paragraph leaks information on the cited cases/paragraphs, which should not be present in a test query \cite{coliee, hou-etal-2025-clerc} as they may lead to inflated performance \cite{Garneau2024NetworkPincites}. Secondly, their document collections are limited only to citing/cited paragraphs, excluding other paragraphs.\footnote{\citet{Olsen2026ProvidingPincite} collected all paragraphs but did not publicly release the dataset.} 
As non-citing and non-cited paragraphs make up the majority of case content, excluding them oversimplifies the task by reducing the pool of candidate paragraphs to search from, creating an unrealistic retrieval setting and overestimating performance \cite{Olsen2026ProvidingPincite}. Upon further inspection of the datasets \cite{Olsen2023ReFraming,Olsen2026ProvidingPincite}, we find extensive data quality and completeness issues (\Cref{sss:issues}), which affect their reusability.

We address the above limitations by introducing \data{}, a new, updated legal IR dataset that can be used for finding relevant legal citations at various query-document levels. \Cref{fig:teaser} overviews the three levels of citation types in our dataset: case-to-case, paragraph-to-case, and paragraph-to-paragraph. \data{} mitigates the data leakage issues in existing datasets by applying heuristics and named entity recognition to remove citation-related information. Our dataset also provides a corpus that includes all case paragraphs, simulating a more realistic retrieval setting. \data{} contributes to advancing legal IR research, with potential extensions that can also benefit adjacent communities, e.g., legal NLP.

\section{Related Work}

\paragraph{Legal IR datasets beyond paragraph-to-paragraph level.} Many legal IR test collections are on the case-to-case or paragraph-to-case levels, e.g., \citet{muser,lecard,coliee,t-y-s-s-etal-2024-ecthr}; \Cref{tab:dataset_comparison} compares existing legal case retrieval or paragraph citation retrieval in English, excluding those with queries that are not derived directly from the case text, e.g., \citet{LegalDivEval2017,upadhya-t-y-s-s-2025-lexclipr,Zheng2025ReasoningBenchmark} as they may lack context specificity. These case- or paragraph-to-case datasets are unusable for the specific task of finding paragraph-level pincites, as they do not pinpoint which specific paragraph is cited; on the other hand, \data{} is designed for this exact purpose, while still allowing case- or paragraph-to-case retrieval.

\paragraph{Legal case retrieval.} Finding relevant legal cases is not only important for establishing precedent and informing judgments, but also for answering questions about the law \cite{Zheng2025ReasoningBenchmark}, e.g., through retrieval-augmented generation \cite{reuter-etal-2025-towards,pipitone2024legalbenchragbenchmarkretrievalaugmentedgeneration,LexRAG2025}. 
Previous work has approached legal case retrieval from many angles \cite{locke2022caselawretrievalproblems,feng-etal-2024-legal}. At query-level, recent work has explored generative rewriting \cite{kim-etal-2025-gure}, knowledge-guided reformulation \cite{deng-etal-2024-learning}, or adding clarification questions \cite{Liu2025CARQ}. Various retrieval strategies also exist, ranging from paragraph-level lexical retrieval \cite{althammer2021dossiercoliee} to domain-specific pre-training based on case structure \cite{SAILER2023,Tang2024CaseGNN}, case connectivity \cite{Tang2024CaseLink}, or inter-case distinctions \cite{Su2025PreTraining}. Other strategies include fine-tuning neural models \cite{Mori2026ICAIL,mahari-etal-2024-lepard}, lexical and semantic reranking \cite{althammer2021dossiercoliee, Askari2024RPRS,hou-etal-2025-clerc}, and incorporating generative legal reasoning into retrieval \cite{li-etal-2026-glier,Tang2026ReaKase,kim-etal-2025-legalsearchlm}.

\section{Dataset}

\subsection{Dataset Construction}
\label{ss:construction}
We construct \data{} by merging two existing datasets \cite{Olsen2023ReFraming,Olsen2026ProvidingPincite}, resolving their quality and reusability issues, and updating them with recent data from the last four years.

\subsubsection{Data Acquisition} We build upon two existing datasets for legal IR: 
\begin{itemize}
    \item \textbf{Unpublished dataset} used in \citet{Olsen2026ProvidingPincite}, which contains all paragraphs (citing, cited, and neither) from CJEU judgments up to 29 July 2024. As the dataset has not been publicly released, we obtain it directly from the authors. The dataset has paragraph-level segmentations and serves as a source of the candidate paragraphs for \data{}.

    \item \textbf{Published dataset} \cite{Olsen2023ReFraming}, which consists of two components: 110,601 pairs of citing and cited paragraphs from CJEU judgments up to 6 October 2021 and 890 relevance annotations by two human experts in law.\footnote{The paragraph pairs can be found at \url{https://huggingface.co/datasets/ngarneau/paragraph_to_paragraph} and all relevance annotations can be accessed at \url{https://github.com/coastalcph/paragraph_network}.} This dataset forms the query and ground truth (relevant) citations for \data{}. 
\end{itemize}
Both datasets originate from EUR-Lex, an online portal that provides official and comprehensive access to EU legal documents.\footnote{\url{https://eur-lex.europa.eu/}} Legal documents published in EUR-Lex may be reused for commercial or non-commercial purposes \cite{eupubsoffice2026legal}, and are often available in multiple EU official languages \cite{linguisticcoverage}.

The datasets (except for the human-annotated paragraphs) provide unique identifiers for each paragraph: CELEX (case ID in EUR-Lex) and the paragraph number. Most paragraph text starts with the paragraph number (see \Cref{fig:teaser}). The datasets also provide case-level metadata, such as the title and date. For more details on their construction and content, we refer the reader to the original papers \cite{Olsen2023ReFraming, Olsen2026ProvidingPincite}.

\subsubsection{Issues and Resolutions} 
\label{sss:issues}
We identify four categories of issues in the two datasets \cite{Olsen2023ReFraming,Olsen2026ProvidingPincite}, which we resolve in \data{}. 

\paragraph{Language ambiguity.} Each dataset contains paragraphs that are either in English or in French, but there is no label indicating their language. This is a problem as different languages may need different processing pipelines (e.g., language-specific model, stemmer, or stopword list). To resolve this, we identify the language with \texttt{langdetect}\footnote{\url{https://pypi.org/project/langdetect/}} and remove non-English paragraphs (about 5\%). 

\paragraph{Missing text.} The citation dataset \cite{Olsen2023ReFraming} is missing the text of 268 unique citing/cited paragraphs (i.e., only having the citing and cited CELEX and paragraph numbers). Missing citing text makes it hard to retrieve relevant information, and missing cited text will not be included in the candidate paragraph collection; both issues may unfairly lower retrieval effectiveness. 
We recover the missing text automatically from the all-paragraph dataset \cite{Olsen2026ProvidingPincite} and also manually by looking them up online on EUR-Lex. During this process, we also correct several errors in the citing/cited paragraph numbers. Yet, some text is not recoverable due to various reasons. Among others, the citing text does not actually have pincites, or the paragraph does not exist in the supposedly cited case (e.g., the case only has 20 paragraphs and the supposedly cited paragraph is number 21). These inconsistencies may be due to erroneous data parsing and information extraction, especially from older documents. Citing-cited paragraph pairs with non-recoverable text are removed.

\paragraph{Segmentation errors.}
The all-paragraph dataset \cite{Olsen2026ProvidingPincite} has missing or incomplete data. We found two types of errors: (i) parsing error (e.g., a pincite is mistaken as the start of a paragraph) affecting more than 1,400 cases; and (ii) consecutive paragraphs are joined (e.g., paragraph number 3 may also contain text from 4). 
To resolve (i), we download the HTML/XHTML version on EUR-Lex using the Cellar API.\footnote{\url{https://op.europa.eu/en/web/cellar/cellar-data/publications}} From the code, we extract the paragraph number and the text of all paragraphs. As paragraph numbers and the text are enclosed within custom tags, the paragraph-level segmentation should be consistent with the source. To resolve (ii), we manually or automatically adjust the segmentation following the text segmentation in the online version (in English or in its original language); if the online versions also have joined paragraphs, we leave them as is.

For consistency between the two datasets that will both be part of \data{}, we ensure that all case CELEX in the citation dataset appear in the all-paragraph dataset. Paragraphs from the missing cases are extracted similarly to the steps above.

\paragraph{Non-reusable human annotations} The human validation data released by  \citet{Olsen2023ReFraming} are not directly usable as the query-paragraph-label triples are not linked back to the paragraph datasets. Further, they are inconsistently formatted (e.g., extra rows/columns, non-standardised labels, mixing Yes/No annotations with notes), rendering them hard to load and use. We manually fix all these formatting issues and link the text back to their ID (CELEX and paragraph number) by performing an exact match on the all-paragraph data. Annotated items that cannot be matched are removed.

\subsubsection{Data Update} To include more recent data, in May 2026, we collected CJEU judgments dated between 1 January 2021--31 December 2025 (inclusive) that are available in English.\footnote{We include a 7-month overlap with the time period of the all-paragraph dataset as a precaution for later-uploaded documents with dates within this period.} Using these criteria, we query EUR-Lex via its advanced search user interface to retrieve a list of case ID (CELEX) and their metadata to be used in the next steps. This results in 2,170 unique cases. For each case, we use \texttt{selenium}\footnote{\url{https://pypi.org/project/selenium/}} to scrape the list of citing and cited case-paragraph pairs\footnote{Under `Instruments cited in case law', e.g., in \url{https://eur-lex.europa.eu/legal-content/EN/ALL/?uri=CELEX:62021CJ0326}} and extract only paragraph-to-paragraph citations to other judgments, resulting in 41,547 unique citations.

To retrieve the text of the citing/cited paragraphs, we extract them from the HTML or XHTML source code using the Cellar API, similar to previous paragraph extraction steps. We then match the text with the citing/cited case-paragraph IDs using a combination of this new data and the all-paragraph dataset. Note that 155 cases are not available online, resulting in their exclusion from our dataset.

\begin{figure*}
    \centering
    \includegraphics[width=0.85\linewidth]{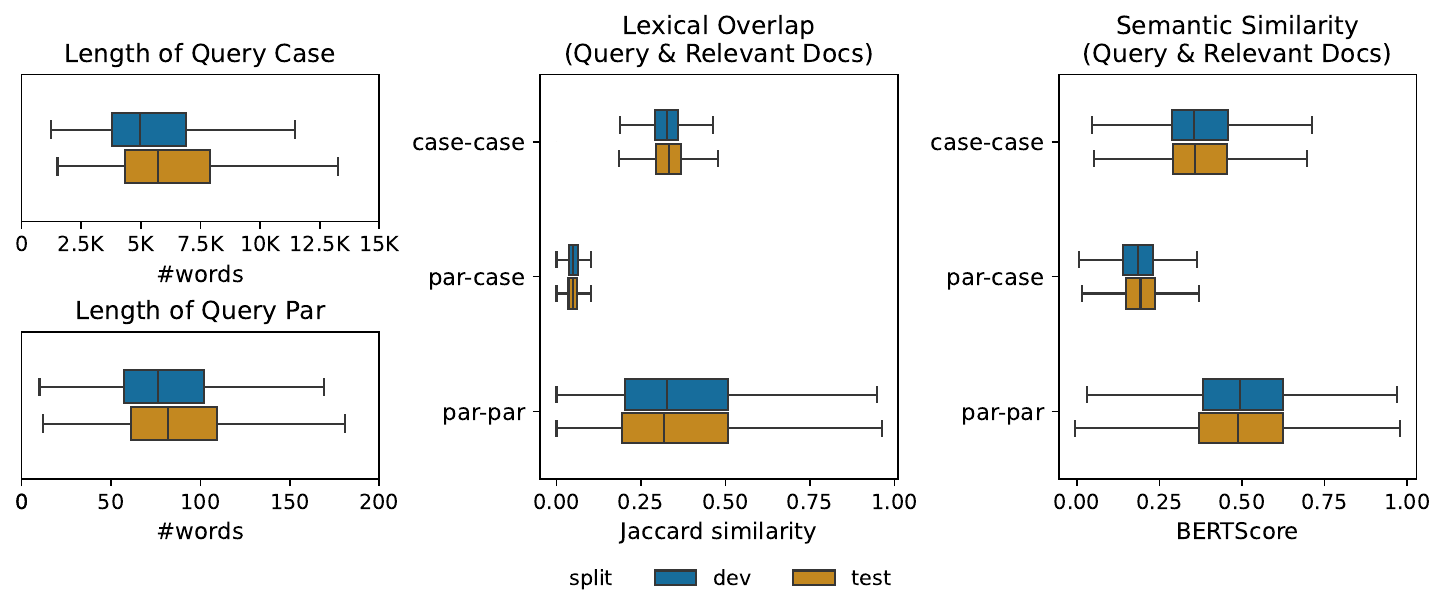}
    \caption{Distribution of query length (left), as well as lexical overlap (centre) and semantic similarity (right) of query and relevant documents.}
    \label{fig:query_qrel_stats}
\end{figure*}

\subsection{Data Preprocessing}
\label{ss:preprocessing}

This section describes how we merge,  clean, and split the dataset, as well as how we mitigate data leakage in the query text.

\paragraph{Data aggregation and cleaning.}
We merge the corrected all-paragraph and the citation datasets with the updated ones and remove duplicates. We then remove trailing whitespaces from the text. To form the case-level query/candidate text, we join the text of all paragraphs in a case. Paragraph-to-case citation data is obtained from treating pincites as citations to a case (see example in \Cref{fig:teaser}), disregarding the cited paragraph number(s). Similarly, we aggregate case-to-case citations from the paragraph-to-paragraph citations.

Regarding the relevance annotations, we aggregate labels across the two experts. For a query-paragraph pair, we consider the paragraph to be relevant to the query if both experts answer `Yes' to at least one of these questions \cite{Olsen2023ReFraming}: 
(i) ``Does the candidate paragraph contain a verbatim version of the rule in the citing paragraph?''; and
(ii) ``Does the candidate paragraph contain a different or more expanded version of the rule in the citing paragraph?''. Non-relevant query results are those that speak about, for example, arguments of the parties (instead of arguments by the court), refer to domestic law (instead of EU law), or speak about the facts of the case (instead of the law of the case). The annotations do not consider non-relevance simply due to a development in the law over time: if it was only that there had been a development in the law, it would not have been coded as not relevant by the expert annotators. 
Pairs that are considered non-relevant are removed; thus, the released annotations only consist of relevant pairs agreed upon by both experts. Before removal of non-relevant pairs, the aggregated labels have an inter-annotator reliability (Cohen's $\kappa$ \cite{Cohen1960}) of 0.42.\footnote{At least one expert answered `Yes' in 72\% of the pairs and both agreed in 84\% of the pairs  \cite{Olsen2023ReFraming}.}

\paragraph{Data split.} We split the aggregated, cleaned data into \textit{train/dev/test} based on year (summarised in \Cref{tab:dataset_split_year}). \textit{Train}, \textit{dev}, and \textit{test} queries are drawn from pre-2018, 2018--2021, and 2022--2025 cases, respectively. For \textit{dev} and \textit{test}, the ground truth (relevant) citations and candidate cases/paragraphs are those with a document year preceding all queries in the split. 
These splits prevent temporal leakage and simulate a realistic legal retrieval setting, e.g., a 2020 case cannot cite a precedent from 2023. While these splits cannot be used to evaluate retrieval at the date of each query, e.g., a 2025 case cannot retrieve a 2023 case, they follow the Cranfield paradigm of having a fixed set of documents \cite{Voorhees2002EvaluationSystems}, similar to the design of common IR test collections.

\begin{table}
\centering
\resizebox{\linewidth}{!}{
\begin{tabular}{lccc}
\toprule
 & train & dev & test \\
\midrule
Query (citing case/paragraph)    &  <2018 &  2018-2021  & 2022-2025  \\
Ground truth (cited case/paragraph)     &  <2018& <2018 &  <2022\\
Candidate case/paragraph & - & <2018 & <2022 \\
\bottomrule
\end{tabular}}
\caption{Year ranges for chronologically splitting query, ground truth, and candidate cases/paragraphs.}
\label{tab:dataset_split_year}
\end{table}

\paragraph{Data leakage mitigation.}
Query text may contain information regarding the citation (e.g., case title, case number, paragraph number, or involved parties). To mitigate data leakage, we employ a pre-trained legal Named Entity Recognition model \cite{kalamkar-etal-2022-named} and regular expressions to identify such information for removal. We analyse the impact of leakage mitigation strategies in \Cref{ss:leakage}.

\subsection{Dataset Statistics}

\begin{table}
    \centering
    \resizebox{\columnwidth}{!}{
    \begin{tabular}{llrrrr}
    \toprule
    query-doc level & split & \#query & \#rel & Avg/q & Max/q \\
    \midrule
    \multirow[c]{3}{*}{case-case} & train & 6,738 & 43,977 & 6.53 & 49 \\
     & dev & 1,263 & 7,799 & 6.17 & 38 \\
     & test & 1,639 & 10,376 & 6.33 & 67 \\
    \cline{1-6}
    \multirow[c]{3}{*}{par-case} & train & 43,847 & 68,394 & 1.56 & 12 \\
     & dev & 9,706 & 12,850 & 1.32 & 12 \\
     & test & 14,504 & 17,659 & 1.22 & 8 \\
    \cline{1-6}
    \multirow[c]{3}{*}{par-par} & train & 43,847 & 85,768 & 1.96 & 94 \\
     & dev & 9,706 & 16,244 & 1.67 & 33 \\
     & test & 14,504 & 23,415 & 1.61 & 88 \\
    \bottomrule
    \end{tabular}}
    \caption{Statistics of query and ground truth for each query-doc level and split. \#rel denotes the number of relevant query-document pairs; Avg/q and Max/q denote the mean and maximum number of relevant documents per query, respectively.}
    \label{tab:main_stat}
\end{table}

Statistics of the final dataset query and ground truth citation (relevance) are presented in \Cref{tab:main_stat}, while corpus statistics and distribution of ground truth annotations based on the source (EUR-Lex, human expert annotations, or both) are shown in \Cref{tab:corpus_stat} and \Cref{tab:qrel_distribution}, respectively. We compute the number of words \texttt{NLTK} tokenizers  \cite{bird-loper-2004-nltk}. Similar to other legal IR corpora, e.g., \cite{coliee}, the documents tend to be long.

\Cref{fig:query_qrel_stats} shows the distribution of query text length, lexical overlap, and semantic similarity with the ground truth (relevant) documents. We compute lexical overlap on the bag of words with Jaccard similarity. Semantic similarity is computed with BERTScore F1 \cite{zhang2019bertscore} using \texttt{distilbert-base-uncased} \cite{Sanh2019DistilBERT} on the first 512 tokens. On average, both lexical overlap and semantic similarity are low (median $<0.5$) across all query-doc levels and splits. The low overlap/similarity indicates that relevance cannot be determined via simple lexical or semantic matching alone \cite{Guo2016DRRM}, making the dataset a challenging and realistic legal IR benchmark.

\begin{table}
    \centering
    \resizebox{\columnwidth}{!}{
    \begin{tabular}{lrr|rr|r}
    \toprule
     & \multicolumn{2}{c|}{number} & \multicolumn{2}{c|}{Avg \#words} & Avg \#par \\
    \midrule
     & case & par & case & par & case \\
    \midrule
    dev & 10,359 & 471,843 & 3,779.21 & 83.00 & 45.55 \\
    test & 12,081 & 593,877 & 4,123.68 & 83.93 & 49.16 \\
    \bottomrule
    \end{tabular}}
    \caption{Statistics of corpus (candidate cases or paragraphs). Avg \#words is the mean number of words per case/paragraph; Avg \#par is the mean number of paragraphs per case.}
    \label{tab:corpus_stat}
\end{table}

\begin{table}
    \centering
    \resizebox{0.9\columnwidth}{!}{
    \begin{tabular}{lrrr}
    \toprule
    level & only EUR-Lex & only human & both \\
    \midrule
    case-case & 7,464 & 246 & 89 \\
    par-case & 12,486 & 286 & 78 \\
    par-par & 15,835 & 331 & 78 \\
    \bottomrule
    \end{tabular}}
    \caption{The number of unique query-doc pairs that are relevant (\textit{qrels}) based on the source (only EUR-Lex/human or both) for various query-doc levels on the \textit{dev} split; \textit{train} and \textit{test} \textit{qrels} are only from EUR-Lex.}
    \label{tab:qrel_distribution}
\end{table}

To illustrate how other candidate documents can serve as strong `distractors', we analyse the similarity between a query and its relevant paragraph (\textit{qrel}) to the similarity between the same query and non-relevant paragraphs (\textit{non-qrel}).\footnote{\textit{Non-qrel} may refer either to documents that are judged not relevant or to unjudged documents.} 
Compu\-ting the similarity for all pairs of queries and paragraphs is impractical, as there are more than 9.7K queries and 471K paragraphs, resulting in 4.6B pairs. Hence, we select a query subset ($n=69$) with ground truth citations from both EUR-Lex and human experts ('EUR-Lex + human' subset); the human validation process had considered highly similar candidate paragraphs that may not be relevant citations, reducing false negatives. 

\begin{figure}
    \centering
    \includegraphics[width=0.7\columnwidth,
    trim=0 0.5cm 0 0.3cm, clip=True
    ]{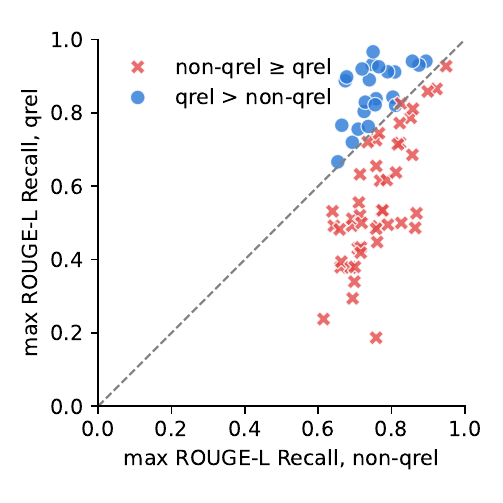}
    \caption{Paired maximum similarity (ROUGE-L recall) per query: query and relevant paragraph (\textit{qrel}) vs.~query and non-relevant paragraph (\textit{non-qrel}). Each point is a query in the `EUR-Lex + human' subset. Red crosses mark instances where the maximum similarity of \textit{non-qrel} exceeds that of the \textit{qrel}, indicating the existence of a strong distractor; blue circles are the opposite.}
    \label{fig:qrel_non_qrel}
\end{figure}

For each query, we compute its order-sensitive lexical similarity to each of the 471K candidate paragraphs, yielding 32.5M pairs.\footnote{The similarity computation took 30 hours.} 
As our order-sensitive lexical similarity, we use ROUGE-L recall \cite{lin-2004-rouge}, a measure based on the longest common subsequence that quantifies how much of the query text is included in the candidate paragraph. \Cref{fig:qrel_non_qrel} shows the maximum ROUGE-L recall of each query against its relevant paragraph versus the maximum ROUGE-L recall of the same query against its non-relevant paragraphs. About 67\% of these queries have a strong distractor, where the highest similarity of the query and non-relevant paragraphs exceeds that of the query and relevant paragraphs, further indicating that a successful retrieval cannot rely on similar order of words alone.

\subsection{Dataset Release}
The release of \data{} follows the FAIR principles \cite{Wilkinson2016TheStewardship}. To facilitate its   \textbf{findability}, \data{} is assigned a DOI (\url{https://doi.org/10.57967/hf/9072}). We publish the dataset on Hugging Face Hub to enhance both its discoverability and \textbf{accessibility}. 
The dataset is also \textbf{interoperable}: it is available in open format (CSV) with widely-used formatting by the IR community (e.g., TREC qrel format) and compatible with common IR experimental frameworks such as   \texttt{pyterrier} \cite{pyterrier}. To facilitate \textbf{reusability}, \data{} is released under the CC-BY license and accompanied with documentation as well as code examples for data loading, retrieval, and reranking.\footnote{\url{https://github.com/theresiavr/legalpincite}} To support future extensions, we include our web scraping and extraction pipelines in the code repository.

\section{Experiments}

\subsection{Experimental Setup}
We describe our setup to evaluate retrieval effectiveness on \data{}. We experiment with all three query-document levels in the dataset, on both the \textit{dev} and \textit{test} splits.

\begin{table*}[t]
    \centering
    \resizebox{0.9\textwidth}{!}{
    \begin{tabular}{cllrrrrrrrr}
 \toprule
level & split & retriever & HR@3 & HR@5 & HR@10 & NDCG@3 & NDCG@5 & NDCG@10 & MAP@1000 & MRR@1000 \\
\midrule
\multirow[c]{8}{*}{\rotatebox[origin=c]{90}{case-case}} & \multirow[c]{4}{*}{dev} & TF-IDF & 0.713 & 0.771 & 0.829 & 0.448 & 0.424 & 0.429 & 0.347 & 0.634 \\
 &  & BM25 & 0.273 & 0.302 & 0.346 & 0.154 & 0.146 & 0.149 & 0.121 & 0.250 \\
 &  & LMIR & \bfseries 0.739 & \bfseries 0.808 & \bfseries 0.873 & \bfseries 0.473 & \bfseries 0.454 & \bfseries 0.462 & \bfseries 0.376 & \bfseries 0.665 \\
 &  & DPH & 0.703 & 0.761 & 0.828 & 0.434 & 0.413 & 0.416 & 0.334 & 0.617 \\
\cline{2-11}
 & \multirow[c]{4}{*}{test} & TF-IDF & 0.719 & 0.793 & 0.849 & 0.447 & 0.425 & 0.427 & 0.337 & 0.644 \\
 &  & BM25 & 0.182 & 0.215 & 0.248 & 0.100 & 0.098 & 0.101 & 0.082 & 0.173 \\
 &  & LMIR & \bfseries 0.728 & \bfseries 0.800 & \bfseries 0.863 & \bfseries 0.456 & \bfseries 0.436 & \bfseries 0.442 & \bfseries 0.353 & \bfseries 0.648 \\
 &  & DPH & 0.693 & 0.759 & 0.821 & 0.421 & 0.399 & 0.403 & 0.316 & 0.613 \\
\cline{1-11}
\multirow[c]{8}{*}{\rotatebox[origin=c]{90}{par-case}} & \multirow[c]{4}{*}{dev} & TF-IDF & \bfseries 0.653 & \bfseries 0.718 & \bfseries 0.792 & \bfseries 0.542 & \bfseries 0.574 & \bfseries 0.604 & \bfseries 0.551 & \bfseries 0.586 \\
 &  & BM25 & 0.248 & 0.280 & 0.318 & 0.198 & 0.213 & 0.227 & 0.206 & 0.222 \\
 &  & LMIR & 0.599 & 0.667 & 0.741 & 0.491 & 0.524 & 0.553 & 0.502 & 0.537 \\
 &  & DPH & 0.560 & 0.622 & 0.697 & 0.458 & 0.488 & 0.516 & 0.469 & 0.502 \\
\cline{2-11}
 & \multirow[c]{4}{*}{test} & TF-IDF & \bfseries 0.647 & \bfseries 0.711 & \bfseries 0.785 & \bfseries 0.545 & \bfseries 0.576 & \bfseries 0.604 & \bfseries 0.553 & \bfseries 0.579 \\
 &  & BM25 & 0.218 & 0.246 & 0.283 & 0.176 & 0.189 & 0.202 & 0.183 & 0.195 \\
 &  & LMIR & 0.597 & 0.664 & 0.737 & 0.498 & 0.529 & 0.557 & 0.507 & 0.534 \\
 &  & DPH & 0.549 & 0.615 & 0.688 & 0.456 & 0.486 & 0.513 & 0.467 & 0.491 \\
\cline{1-11}
\multirow[c]{8}{*}{\rotatebox[origin=c]{90}{par-par}} & \multirow[c]{4}{*}{dev} & TF-IDF & 0.652 & 0.715 & 0.783 & \bfseries 0.520 & 0.546 & 0.573 & 0.519 & 0.582 \\
 &  & BM25 & \bfseries 0.654 & \bfseries 0.719 & \bfseries 0.785 & \bfseries 0.520 & \bfseries 0.548 & \bfseries 0.574 & \bfseries 0.520 & \bfseries 0.583 \\
 &  & LMIR & 0.633 & 0.702 & 0.770 & 0.500 & 0.528 & 0.555 & 0.500 & 0.563 \\
 &  & DPH & 0.621 & 0.688 & 0.758 & 0.489 & 0.516 & 0.543 & 0.489 & 0.549 \\
\cline{2-11}
 & \multirow[c]{4}{*}{test} & TF-IDF & 0.628 & 0.692 & 0.766 & 0.502 & \bfseries 0.529 & 0.555 & 0.502 & 0.559 \\
 &  & BM25 & \bfseries 0.631 & \bfseries 0.694 & \bfseries 0.769 & \bfseries 0.503 & \bfseries 0.529 & \bfseries 0.557 & \bfseries 0.503 & \bfseries 0.560 \\
 &  & LMIR & 0.607 & 0.676 & 0.754 & 0.481 & 0.509 & 0.536 & 0.482 & 0.538 \\
 &  & DPH & 0.601 & 0.663 & 0.738 & 0.475 & 0.501 & 0.528 & 0.475 & 0.530 \\
\bottomrule
\end{tabular}
}
\caption{Baseline effectiveness across query-document levels and splits. The best scores per level/split are bolded.}
\label{tab:main_result}
\end{table*}

\paragraph{Baselines.} We employ four bag-of-words retrieval baselines. Following \citet{muser,lecard}, we select TF-IDF \cite{SALTON1988513}, BM25 \cite{bm25}, and LMIR (based on Dirichlet language model, \citet{dirichletLM}), in addition to DPH \cite{DPH}, a parameter-free, hypergeometric model. All baselines are run with \texttt{pyterrier} \cite{pyterrier}. Except for the parameter-free DPH, we tune the hyperparameters of all baselines on the \textit{dev} split, using grid search following the range used in prior work \cite{Askari2021CaseLaw,dirichletLM}.\footnote{We include the search space and best configurations in the code repository.} 
The configuration with the best NDCG@10 is then used for testing.

\paragraph{Neural rerankers.} Retrieval performance can be enhanced through semantic matching \cite{Guo2016SemanticMatching} and contextualised representations \cite{nogueira2020passagererankingbert}. 
To this end, we use three zero-shot rerankers that are based on: 
(i) LegalBERT, which is pre-trained on EUR-Lex text \cite{chalkidis-etal-2020-legal};
(ii) SAILER \cite{SAILER2023}, a structure-aware pre-trained legal case retrieval model that is fine-tuned on COLIEE {\color{blue}
\cite{COLIEE2020,COLIEE2021}}; and
(iii) BGE (\texttt{bge-reranker-v2-m3}, \citet{chen2024bge}). Altogether, the rerankers include pre-trained language models fine-tuned in the legal domain and a general-purpose model (i.e., BGE), as well as both bi- and cross-encoder architectures: LegalBERT and SAILER are used as bi-encoders, scoring relevance via the cosine similarity between independently computed query and document representations. While SAILER pools the \texttt{[CLS]} token from the last hidden layer \cite{SAILER2023}, for LegalBERT we use mean pooling over all tokens. In contrast, BGE is a cross-encoder that jointly encodes each query-document pair and predicts a relevance score directly via a classification head. All inputs are truncated to a maximum of 512 tokens \cite{Su2025PreTraining}; this limit is applied to the query and document independently for the bi-encoders, but jointly to the concatenated pair for the cross-encoder.  For a case-level query or document, the reranker may therefore only consider a small fraction of the text.

\paragraph{Evaluation.} We measure retrieval effectiveness at cut-off $k\in \{3, 5, 10\}$ with Hit Rate (HR@$k$) and NDCG@$k$ \cite{Jarvelin2002CumulatedTechniques}. We also compute Mean Average Precision (MAP) and Mean Reciprocal Rank (MRR) at top-1000.

\subsection{Multi-level Retrieval}

\paragraph{Retrieval performance.} \Cref{tab:main_result} presents the effectiveness of baseline retrievers on case-to-case, paragraph-to-case, and paragraph-to-paragraph levels for both \textit{dev} and \textit{test} splits. Generally, effectiveness does not differ much between the two splits, but significantly differs across the retrieval levels: LMIR performs best for case-to-case, while TF-IDF performs best for paragraph-to-case. For paragraph-to-paragraph, BM25 scores the best, on par with TF-IDF. DPH maintains a stable performance across all splits and levels. 
We note that BM25 has the least stable performance across different levels, scoring considerably worse for case-to-case and par-to-case. 

\begin{table}
\centering

\resizebox{\columnwidth}{!}{
\begin{tabular}{llrrrr}
\toprule
level & subset & TF-IDF & BM25 & LMIR & DPH  \\
\midrule
\multirow[c]{2}{*}{case-case} & EUR-Lex only & 0.428 & 0.147 & 0.462 & 0.415 \\
 & EUR-Lex + human & 0.447 & 0.190 & 0.475 & 0.456 \\
\cline{1-6}
\multirow[c]{2}{*}{par-case} & EUR-Lex only & 0.604 & 0.227 & 0.553 & 0.516 \\
 & EUR-Lex + human & 0.594 & 0.172 & 0.541 & 0.511 \\
\cline{1-6}
\multirow[c]{2}{*}{par-par} & EUR-Lex only & 0.573 & 0.574 & 0.554 & 0.542 \\
 & EUR-Lex + human & 0.625 & 0.628 & 0.620 & 0.603 \\
\bottomrule
\end{tabular}}
\caption{Comparison of NDCG@10 for two disjoint subsets of the \textit{dev} split: queries with ground truth sourced from EUR-Lex (EUR-Lex only) and queries with additional human expert validation (EUR-Lex + human).
}
\label{tab:hj_non_hj}
\end{table}

\paragraph{Ground truth source comparison} To compare the difference between EUR-Lex and additional human relevance annotations in the retrieval evaluation, we report in \Cref{tab:hj_non_hj} the retrieval performance (NDCG@10) on the \textit{dev} split for queries whose ground truth citations are sourced only from EUR-Lex (denoted by `EUR-Lex only') and the remaining queries, i.e., those with additional human validation (EUR-Lex + human). Except for BM25, which scores low for both query subsets in case-case and par-case, in general, the difference between the two is relatively minor,
indicating that evaluation on the much smaller EUR-Lex + human query subset can reasonably estimate those obtained on the full dataset. Consequently, to save computation, one can consider evaluating only on the `EUR-Lex + human' query subset.

\paragraph{Semantic reranking.} 

We investigate the potential benefit of semantic reranking in a retrieve-then-rerank pipeline: for each query in the `EUR-Lex + human' subset, we rerank the top-50 documents retrieved by TF-IDF, which performs relatively well across all query-document levels. 
\Cref{tab:semantic_reranking} shows the reranking performance. Across all levels, none of the rerankers outperform the first-stage retriever; this is in line with recent findings that lexical retrievers could outperform zero-shot dense models \cite{Mori2026ICAIL} and that semantic reranking may degrade legal retrieval performance \cite{hou-etal-2025-clerc}. Interestingly, the general-purpose BGE reranker outperforms the domain-specific rerankers for par-case and par-par; this is similar to \citet{SAILER2023}, who found that LegalBERT can perform worse than BERT for legal retrieval. Overall, our findings suggest that in these constrained experimental settings, lexical retrievers could outperform zero-shot semantic rerankers for legal IR.

\begin{table}
    \centering
    \resizebox{\columnwidth}{!}{
    \begin{tabular}{llll}
    \toprule
     & case-case & par-case & par-par \\
    \midrule
    TF-IDF & \textbf{0.447} & \textbf{0.594} & \textbf{0.625} \\
    \quad +LegalBERT \cite{chalkidis-etal-2020-legal} & 0.307 & 0.239 & 0.456 \\
    \quad +SAILER \cite{SAILER2023} & 0.312 & 0.205 & 0.430 \\
    \quad +BGE \cite{chen2024bge} & 0.262 & 0.271 & 0.583 \\
    \bottomrule
    \end{tabular}
    }
    \caption{Semantic reranker performance (NDCG@10) on the top-50 retrieved results of TF-IDF on the EUR-Lex + human query subset (\textit{dev} split), across query-document levels. Bold indicates the best score per level.}
    \label{tab:semantic_reranking}
\end{table}

\subsection{Data Leakage Mitigation Analysis}
\label{ss:leakage}

\citet{Garneau2024NetworkPincites,Olsen2026ProvidingPincite} found that identifiers (IDs) at the start of paragraphs inflate retrieval performance and mitigated it by removing them. 
We compare the impact of data leakage mitigation by evaluating retrieval performance (NDCG@10) for all queries on the test split under three settings: 
(i) \textit{ori}: original (unmasked) query and document text;
(ii) \textit{w/o par ID}: removal of ID from the start of paragraphs, both in queries and documents;
(iii) \textit{w/o citation}: removal of cited case/paragraph information from queries (see \Cref{ss:preprocessing}).

\begin{table}
\centering
\resizebox{\columnwidth}{!}{
\begin{tabular}{llllll}
\toprule
level & setting &
TF-IDF &
BM25  &
LMIR &
DPH \\
\midrule
\multirow{3}{*}{case-case} & ori & 0.429 (+0.5\%) & 0.070 (-30.7\%) & 0.447 (+1.1\%) & 0.411 (+2.0\%) \\
 & w/o par ID & 0.429 (+0.5\%) & 0.137 (+35.6\%) & 0.447 (+1.1\%) & 0.415 (+3.0\%) \\
 & w/o citation & 0.427  & 0.101  & 0.442  & 0.403  \\
\cline{1-6}
\multirow{3}{*}{par-case} & ori & 0.622 (+3.0\%) & 0.255 (+26.2\%) & 0.657 (+18.0\%) & 0.611 (+19.1\%) \\
 & w/o par ID & 0.604 (+0.0\%) & 0.313 (+55.0\%) & 0.609 (+9.3\%) & 0.608 (+18.5\%) \\
 & w/o citation & 0.604  & 0.202  & 0.557  & 0.513  \\
\cline{1-6}
\multirow{3}{*}{par-par} & ori & 0.463 (-16.6\%) & 0.462 (-17.1\%) & 0.429 (-20.0\%) & 0.446 (-15.5\%) \\
 & w/o par ID & 0.436 (-21.4\%) & 0.430 (-22.8\%) & 0.350 (-34.7\%) & 0.421 (-20.3\%) \\
 & w/o citation & 0.555  & 0.557  & 0.536  & 0.528  \\
 \bottomrule
\end{tabular}}
\caption{Comparison of test NDCG@10 under various leakage mitigation settings: \textit{ori} uses the original (unmasked) queries and corpus, \textit{w/o par ID} removes paragraph identifiers from both, and \textit{w/o citation} removes citation information from queries. Percentages denote relative change w.r.t.~\textit{w/o citation}.
}
\label{tab:leakage}
\end{table}
\Cref{tab:leakage} presents both the raw NDCG@10 scores and their change relative to the \textit{w/o citation} setting. Generally, for case-to-case and paragraph-to-case, the citation-related data leakage induced by \textit{ori} and \textit{w/o par ID} increases NDCG; the increase is relatively marginal, up to 0.036 (35.6\%) for case-to-case, while the change for paragraph-to-case is bigger, up to 0.111 (55.0\%). An exception to this is BM25 on case-to-case, where NDCG drops by 0.031 (30.7\%) in \textit{ori}. 
Notably, \textit{w/o citation} outperforms other settings for paragraph-to-paragraph, with up to 0.186 NDCG@10 difference (34.7\%). We posit that removing citation information also reduces noise in the query \cite{mahari-etal-2024-lepard}, thereby improving retrieval. This may also be why \textit{w/o par ID} sometimes outperforms the \textit{ori} setting.

Overall, we show that simply removing paragraph IDs to mitigate leakage is insufficient, as it may still inflate performance. We therefore recommend using the queries provided in \data{}, in which citation-related information has been filtered.

\section{Conclusion}
We present \data{}, a large-scale and multi-level legal IR test collection in English that can be used to do three types of retrieval: finding case paragraphs to be cited in a paragraph (paragraph-to-paragraph); and finding cases that are relevant to a query case (case-to-case) or to a query paragraph (paragraph-to-case).
Beyond legal IR, \data{} can potentially be extended for other tasks, e.g., citation link prediction, cross-lingual retrieval (given its multilingual origin), legal textual entailment (by extracting relevant portions of cited paragraphs), legal Retrieval-Augmented Generation (by pairing preliminary-ruling questions with their answers), and evaluation of Large Language Model legal reasoning.

\section*{Limitations} We acknowledge several limitations of \data{}. The ground truth citations are mostly sourced from EUR-Lex and may suffer from feedback loop: judges might have searched EUR-Lex (or other legal IR systems) for relevant citations and cited the top-retrieved results. The text may contain verbatim copies of the cited paragraph, but this not a dominant pattern: the lexical and semantic overlap between queries and their relevant documents (\Cref{fig:query_qrel_stats}) is generally low (median < 0.5). Moreover, some non-relevant paragraphs' lexical similarity to the query is higher than the relevant paragraph's similarity to the same query (\Cref{fig:qrel_non_qrel}). Further, such text could also originate from the judgment author's experience and domain knowledge.

The expert annotations were done on the top-10 results of a dense retriever \cite{Olsen2023ReFraming}, which may limit their exhaustiveness. Yet, this aligns with legal citation practices where citing a few relevant sources suffices \cite{Olsen2023ReFraming,Derlen_Lindholm_2015}. As such, evaluation should focus on precision-oriented metrics at short cut-offs rather than recall. We note that annotating legal text requires domain expertise, which means that the task cannot be crowdsourced, and that such annotation is time-consuming due to lengthy text.

Additionally, we have employed an automatic pipeline to remove non-English text and mask citation information in queries. While some text (especially older cases) may still be in French, we did not find any non-English text in a random sample of 300 paragraphs (100 each from the \textit{dev} queries, the \textit{test} queries, and the \textit{test} corpus). As for the citation masking performance, we have shown in \Cref{tab:leakage} that it has a notable impact on retrieval effectiveness. As such, while there may be missed instances, we do not expect them to significantly impact performance, and a more systematic audit of such removals, e.g., at span level, is left for future work.

\section*{Ethical Considerations}  Judgments may contain names of people and entities. While we have tried to remove them as part of the citation masking process, some information may remain. Nevertheless, the content of judgments is also available publicly on EUR-Lex.

\section*{Acknowledgments}
We thank Yannis Panagis for creating the two original datasets \cite{Olsen2023ReFraming,Olsen2026ProvidingPincite} that served as the foundation for this work.

\bibliography{references}

\end{document}